# Asymmetric Data Acquisition System for an Endoscopic PET-US Detector

Carlos Zorraquino, Ricardo Bugalho, Manuel Rolo, Jose C. Silva, Viesturs Vecklans, Rui Silva, Catarina Ortigão, Jorge A. Neves, Stefaan Tavernier, Pedro Guerra, Andres Santos and João Varela

*Abstract*—According to current prognosis studies of pancreatic cancer, survival rate nowadays is still as low as 6% mainly due to late detections. Taking into account the location of the disease within the body and making use of the level of miniaturization in radiation detectors that can be achieved at the present time, EndoTOFPET-US collaboration aims at the development of a multimodal imaging technique for endoscopic pancreas exams that combines the benefits of high resolution metabolic information from Time-Of- Flight (TOF) Positron Emission Tomography (PET) with anatomical information from ultrasound (US). A system with such capabilities calls for an application-specific high-performance Data Acquisition System (DAQ) able to control and readout data from different detectors. The system is composed of two novel detectors: a PET head extension for a commercial US endoscope placed internally close to the Region-Of-Interest (ROI) and a PET plate placed over the patient's abdomen in coincidence with the PET head. These two detectors will send asymmetric data streams that need to be handled by the DAQ system. The approach chosen to cope with these needs goes through the implementation of a DAQ capable of performing multi-level triggering and which is distributed across two different on-detector electronics and the off-detector electronics placed inside the reconstruction workstation.

This manuscript provides an overview on the design of this innovative DAQ system and, based on results obtained by means of final prototypes of the two detectors and DAQ, we conclude that a distributed multi-level triggering DAQ system is suitable for endoscopic PET detectors and it shows potential for its application in different scenarios with asymmetric sources of data.

## I. Introduction

MOTIVATED by the high mortality rates for pancreatic cancer, the location of the organ under study (close to an open cavity) and current possibilities on detector radiation miniaturization, the EndoTOFPET-US collaboration [1] was conceived to create an endoscopic PET scanner in order to provide to physicians with a tool to study newest specific biomarkers for pancreatic and prostatic cancer. This novel scanner aims to push current limitations of whole body PET detectors, making possible detections of millimetric lesions. The standout characteristic of this system is its Region-Of-Interest (ROI) specific configuration, which breaks with the traditional whole body scheme thanks to the incorporation of a PET detector head in an endoscopic probe (see Fig. 1).

Similar US endoscopic systems in combination with PET scanners are under investigation [2], nevertheless

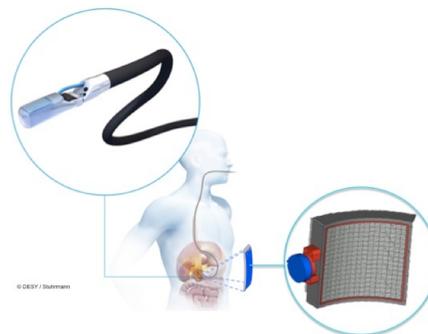

Fig. 1. The EndoTOFPET-US scanner composed of an abdominal PET plate in coincidence with a miniaturized PET extension head on the tip of an ultrasound probe which is placed close to the ROI (in the duodenum in this case which is the closest accessible region with respect to the pancreas).

ENDOTOFPET-US is the first scanner that miniaturizes a PET detector to introduced it inside the patient body to cope with the image quality requirements of the application.

In order to obtain the intended image quality for early detections, the system user requirements cover several technological challenges: 1 mm image spatial resolution, unprecedented 200 ps Coincidence Time Resolution (CTR) for enhanced background rejection, online tracking of both detectors and image reconstruction with partial volume information from an asymmetric geometry [3-4].

A system with such characteristics demands a high-performance application-specific Data Acquisition (DAQ) system able to configure, control and readout data simultaneously from two different detectors.

## II. System Architecture

The EndoTOFPET-US DAQ [5-6], can be decomposed in three different subsystems attending to their functionalities and location, as it is seen in Fig. 2:

Manuscript received June 16, 2014. This work, as part of PicoSEC MCNet Project, is supported by a Marie Curie Early Initial Training Network Fellowship of the European Community's Seventh Framework Programme under contract number (PITN-GA-2011-289355-PicoSEC-MCNet). And EndoTOFPETUS has received funding from the European Union 7[th] Framework Program (FP7/2007-2013) under Grant Agreement No. 256984.

Carlos Zorraquino, Ricardo Bugalho, Manuel Rolo, Jose Carlos Silva, Viesturs Vecakalns, Rui Silva, Catarina Ortigão, Jorge Neves and João Varela are with the Laboratório de Instrumentação e Física Experimental de Partículas, Lisboa, 1000-149 PT (e-mail: carloszg@lip.pt).

Pedro Guerra and Andrés Santos are with the Biomedical Image Technologies Lab and the CIBER-BBN, Universidad Politécnica de Madrid, Madrid, 28040 SP.

Stefaan Tavernier is with the Vrije Universiteit Brussel, Brussel, 1050 BE.

1. On-detector electronics at the external plate, this submodule implements the segment of the DAQ contained in the abdominal PET plate;
2. On-detector electronics at the internal probe, this submodule corresponds to the segment of the DAQ contained in the PET extension of the endoscopic probe;
3. Off-detector electronics, which refers to the segment of the DAQ contained within the image reconstruction workstation.

Thanks to the approach of distributing the DAQ over three subsystems, global complexity is balanced and it provides the possibility to perform and manage centrally a multi-level triggering scheme.

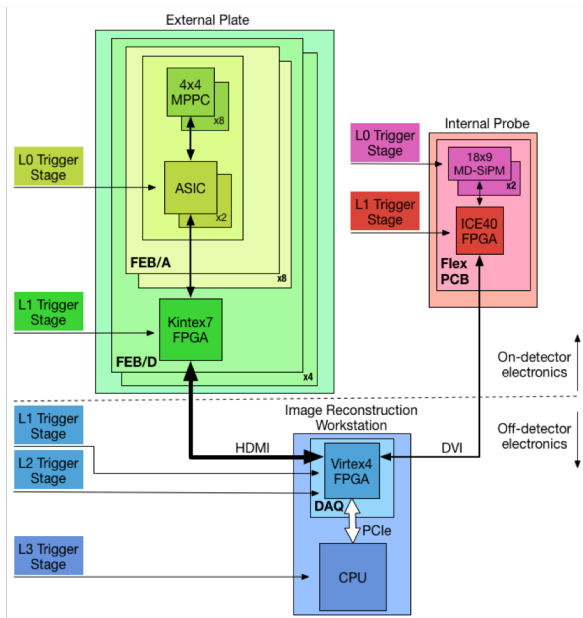

Fig. 2. DAQ system architecture. Top left picture shows the DAQ on-detector external plate electronics making emphasis on the different components, its arrangement and triggering stages. On a similar way, same characteristics are depicted for the on-detector internal probe electronics in the top right of the figure and off-detector electronics in the bottom.

Such a triggering system gives the freedom to apply different principles on each triggering stage:

- Level 0 triggering stage. At this stage gamma events signals are discriminated as valid or not according to the event energy, rejecting the dark counts produced by the photo-detector. This stage is implemented within the ASIC at the external plate as well as in the md-SiPM at the internal probe. The threshold values are adjusted from the image reconstruction workstation.
- Level 1 triggering stage. A symmetric error handling mechanism has been implemented in order to preserve detector uniformity. At this stage, if a frame coming from a certain detector and containing the gamma events detected within a concrete time lapse is lost, then all the incoming gamma events from the other sources within the same time window have to be discarded in order to avoid introducing artifacts in the image reconstruction algorithm.
- Level 2 triggering stage. It is an early and coarse stage of temporal coincidence classification and it is performed online in the off-detector DAQ FPGA, classifying as temporal coincident all events lying within the same 12.5 ns time window.
- Level 3 triggering stage. This is the fine stage of temporal coincidence classification and it is performed later on by the image reconstruction workstation software.

The direct benefit of this triggering scheme is that data volume is progressively reduced attending to the information available at every stage. Therefore, the required bandwidth (BW) in the interfaces between the different system components is considerably reduced.

The proposed system topology is highly asymmetric due to the geometrical differences of the two different detectors (plate and probe) as it can be observed on the detectors definition exposed within the next section. Consequently, the detectors readout has to face completely unbalanced data input streams and thus specific needs are required in terms of gamma events buffering, temporal sorting and temporal coincidence classification.

Specific information on the PET data transmission defined for this system is available in [5], where PET packets' content and format are described and expected gamma events rates are studied for both detectors (40MHz for the external plate and 200kHz for the internal probe).

### III. SYSTEM COMPONENTS

#### A. External Plate

*1) External Plate Definition and Characteristics*

The EndoTOFPET-US external plate is a pixelated PET detector panel placed over the patient's abdomen. It is composed of a total of 256 matrices, with 4x4 arrays of LYSO:Ce scintillating crystals per matrix. Each of them is coupled to a coupled to a Hamamatsu Through Silicon Via – Multi Pixel Photon Counter (TSV-MPPC) array of 4x4 Silicon Photomultiplier (SiPM) pixels.

As it was already introduced in Fig. 2, a modular approach has been chosen to implement the plate electronics leading to the division of the panel into four independent submodules.

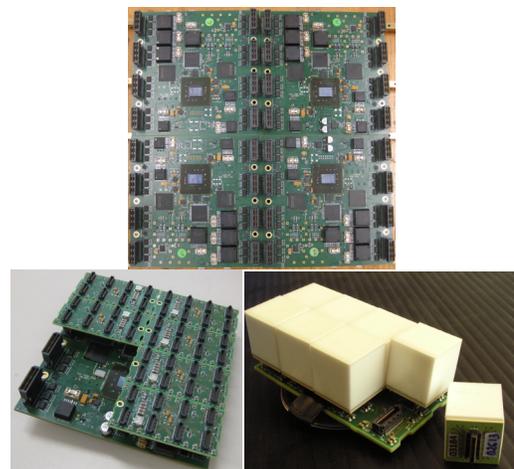

Fig. 3. External plate detector. Top picture shows the 4 FEB/Ds that conform the external plate, the bottom left one shows the assembly of the 8 FEB/As on a FEB/D and the bottom right picture shows how the 8 crystal matrices with their MPPCs glued are connected in a FEB/A.

Each of the 4 quarters contains a Front-End-Board (FEB) called FEB/D (D, because it interfaces with the DAQ), which in turn are connected to 8 FEB/A each (A, because it contains two ASICs). As a result, each FEB/A interfaces 8 MPPC 4x4 arrays and thus the FEB/D FPGA (a Kintex 7 from Xilinx San Jose, USA) reads a total of 1024 MPPC channels, see Fig. 3. The height of the FEB/D to FEB/A connectors allow the insertion of a cooling plate between the two boards.

1.1) *FEB/A*

Fig. 4 depicts an overview of the main components of a FEB/A:

- 8 LYSO:Ce scintillating crystal matrices
- 8 MPPCs
- 2 ASICs [8-9]
- 2 Fan out chips

Each FEB/A contains 8 LYSO:Ce scintillating crystal matrices with 4x4 pixels from Crystal Photonics, Inc. Each matrix is composed of 16 3.5x3.5x15 mm3 pixels, separated by a thin reflective layer from Vikuiti$^{TM}$. Details on the characterization of these matrices including light yield (LY) and temporal response measurements can be found in [7].

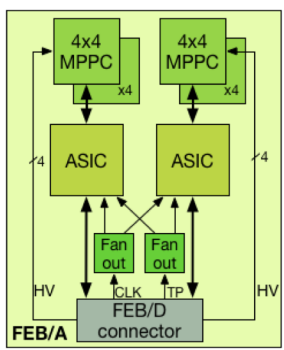

Fig. 4. FEB/A components. Block diagram showing the most relevant FEB/A components from the DAQ point of view.

The 8 MPPC 4x4 arrays (Hamamatsu TSV-MPPC S12643-050CN) glued to the crystal matrices are biased through a high-voltage (HV on Fig. 4) dedicated line from the FEB/D, which tunes independently every photo-detector 4x4 array. The biasing fine-tuning of each of the 16 MPPC channels is performed by adjustment of the voltage base line on every ASIC channel.

MPPC arrays readout is performed by means of a low power, low noise and high bandwidth ASIC with 64-channels implemented by the consortium [8-9]. The ASIC is able to trigger on the first photoelectron, while rejecting dark counts (L0-Plate Trigger Stage), at a maximum rate of 160 Kevents/s per channel. The ASIC digital interface encapsulates all the accepted gamma events in a digital frame. Each frame assembles events captured through its 64 channels within 6.4 μs time windows. For each event the ASIC includes energy and time information. For EndoTOFPET-US application, where the Field-of-View (FoV) is small and detector sensitivity is expected to be low, special emphasis has been placed on the optimization of timing resolution with the goal of improving the quality of the final image. For this reason, the ASIC has been designed to be able to provide gamma events fine time information with a 50 ps time binning.

Sensitivity simulations with the software package GATE v6 predict a value around 10 cps/kBq [10].

Given the risk of using a novel technology for MPPCs readout, two different ASICs have been independently implemented by the collaboration: TOFPET ASIC [8] and STiC ASIC [9]. The STiC and TOFPET ASICs will be used to assemble two versions of the external plate, dedicated to prostate and pancreatic cancer detection respectively. The DAQ has been designed to be flexible in the processing of the ASIC data for the two different models, FEB/A is chip specific and FEB/D is able to accept FEB/A_TOFPET or FEB/A_STiC boards by selecting one of the two firmware versions designed to interface each ASIC.

These two ASICs have been already characterized showing similar performance. The energy resolution for the 511 keV photo-peak is better than 20% and the coincident time resolution (CTR) using LYSO crystals and MPPC photo-detectors is in the range 200-300 ps depending on the measurement conditions. Further details on the two ASICs characterization can be found in previous publications [11-12].

The communication protocol implemented between the ASIC and the FEB/D FPGA provides point-to-point serial communication at a maximum data rate ranging from 160 up to 640 Mb/s depending on the configuration used. It is based on 8B/10B codification and it includes extra error detection mechanisms.

The ASICs reference clock signals (CLK on Fig. 4), which are working at 160 MHz, and the test-pulse signals (TP on Fig. 4) used to electrically trigger a gamma event on the ASIC for calibration purposes, are distributed through 1:2 low jitter LVDS buffers (CLVD2102, Texas Instruments, Texas, USA).

1.2) *FEB/D*

Fig. 5 shows an overview of the main components of a FEB/D:

- 8 FEB/As
- 2 HV DAC
- Fan out chips
- Clock synthesizer
- Serial flash memory
- Local oscillator
- 2 HDMI interfaces
- FPGA

The two HV Digital to Analog Converters (DAC) (AD5535BKBCZ, Analog Devices, Norwood USA) are configured by the Kintex7 (Xilinx, San Jose, USA) FPGA to provide the high-voltage biasing to 32 MPPC matrices each, which can be configured with a granularity of 6.25 mV.

The precision clock synthesizer / jitter attenuator (SI5368B-C-GQ, Silicon Laboratories, Austin, USA) is configured by the Kilinx7 FPGA in order to generate three reference clock signals. All the generated reference clock signals are synchronized to the external reference clock coming from the off-detector DAQ card through the HDMI/m interface. Having a common external clock reference for both detectors is a must in order to have coherent time measurements. The three generated reference clocks are: the ASIC reference clock sent to the FEB/A (160MHz and 640MHz for TOFPET and STiC ASICs respectively), a replica of this as system reference clock (Syst ref on Fig. 5) and the FPGA Multi-Gigabit Transceivers (MGT) reference clock. The FPGA MGTs are used to transmit and receive data through the HDMI interfaces.

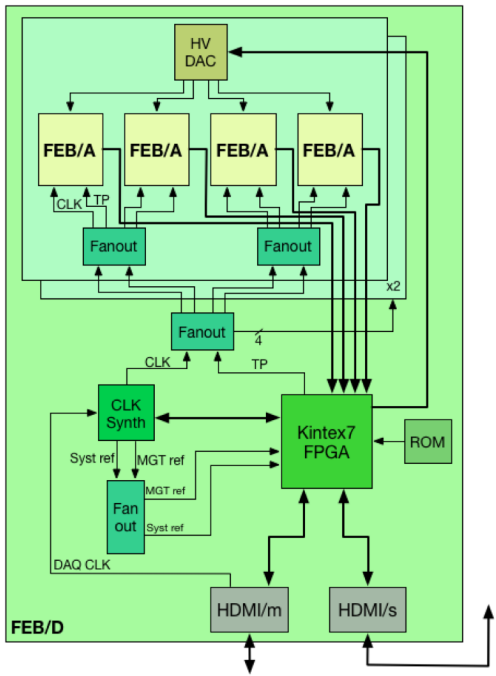

Fig. 5. FEB/D components. Block diagram showing the most relevant FEB/D components from the DAQ point of view.

In the FEB/D the reference clock and test-pulse signals are distributed through low jitter LVDS buffers towards the FEB/As. In this case, the devices used are Texas Instruments CDCLVD2104RHDR (dual 1:4 fan-out) and CDCLVD2102RGTR (dual 1:2 fan-out).

During power-off the Kintex7 FPGA firmware is stored in a serial flash memory (ROM in Fig 5), Micron N25Q128A13ESE40F. On FEB/D boot-up, firmware is loaded from this memory into the FPGA.

The communication protocol chosen between the FEB/D and the off-detector DAQ card is based on the AURORA 8B/10B protocol (Xilinx, 2100 Logic Drive San Jose, CA) working at 1.6 Gb/s over a HDMI physical link. The implementation includes extra error detection mechanisms (L1-Plate Trigger Stage). The overall resulting event transmission rate goes up to 64/128 Mevents/s for full/compact event format respectively.

One last import remark about the On-detector DAQ FEB/D electronics is that it has been designed under the premise of flexibility and reusability. Therefore, the FEB/D firmware accepts two different system topologies for system integration ease: either each FEB/D is connected independently to the DAQ or FEB/Ds are paired in a daisy chain configuration reducing the number of physical links between the on-detector plate electronics and the off-detector DAQ from four down to two HDMI cables. In a similar way, for future applications of this DAQ, more than two FEB/Ds can be daisy-chained to increase the number of channels in the detector. As it can be observed in Fig. 5, FEB/D has two micro-HDMI connectors: HDMI/m (master) for the connection towards the off-detector DAQ on the chain and HDMI/s (slave) for the connection away from the off-detector DAQ on the chain.

2) *External Plate Functionality*

The FEB/D FPGA is responsible for the implementation of the functionalities needed for the DAQ segment tied to the On-detector plate electronics:

1. Gamma events routing and concentration. Readout of 16 ASICs centralizing into a super-frame all gamma events detected on the associated 1024 MPPC channels during a 6.4 µs time window and its transmission to the off-detector DAQ electronics using MGTs;
2. System configuration. Delivering configuration commands from the off-detector DAQ to the different devices on the FEB/D and FEB/As (namely ASICs, HV-DAC and CLK synthesizer);
3. System monitoring. Readout of FPGA internal parameters and FEB/D and FEB/As parameters to assess system performance online.

3) *External Plate Validation*

The external plate electronics and their corresponding segment of the DAQ have been validated and characterized with the experimental setup shown in Fig. 6. This setup uses the following modules:

1. The final prototype of the off-detector DAQ card;
2. A TOFPET ASIC test board implemented for the characterization of this chip;
3. A ML605 Xilinx Virtex6 development kit including a custom made mezzanine board that provides 4 micro-HDMI IOs through the kit FMC connector.

Connecting boards 2 and 3 by a FMC2FMC flex connector together they conform a FEB/D + 1xFEB/A equivalent capable of validating the configuration and readout of the two ASICs present on the FEB/A with its associated 8 MPPC 4x4 arrays (readout of 128 MPPCs channels).

The reason for using a development kit instead of real FEB/D + FEB/As prototypes to conduct these measurements is that, after ASIC production, ASIC characterization on TOFPET ASIC test board is needed for an optimum design of FEB/D and FEB/A boards. Therefore, we had an available experimental setup for DAQ validation before FEB/D and FEB/As manufacturing, which is used for ASIC characterization and DAQ validation in parallel.

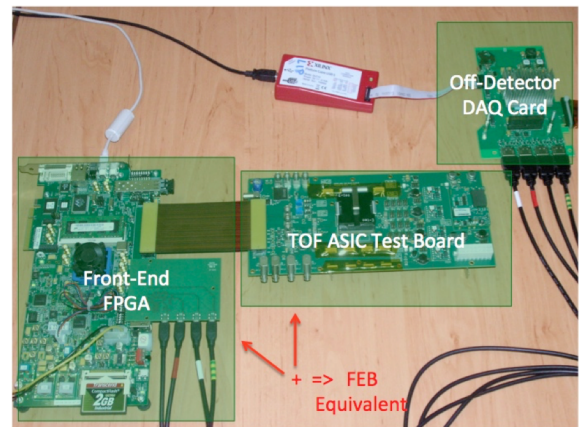

Fig. 6. External plate test setup. Top right corner of the figure shows the final prototype of the off-detector DAQ card making use of the 4 HDMI IOs simultaneously to communicate with the FEB/D + 1xFEB/A equivalent conformed by the Xilinx Virtex6 development kit shown in the bottom left of the picture and the TOFPET ASIC test board in the center of the picture.

Off-detector DAQ electronics accepts up to four high-speed links through HDMI connections. Experimental tests making use of the four present links transmitting simultaneously from the on-detector to the off-detector electronics, demonstrate that 1.6 Gb/s can be used simultaneously on each link maintaining an error free data transmission during 24 hours continuous acquisitions. For the quality measure of this link, a specific firmware version has been implemented for the Virtex6 where 4 MGTs are instantiated to externally loopback the information transmitted by the four MGTs present on the off-detector DAQ card where (Bit Error Rate) BER is computed.

*B. Internal Probe*

1) *Internal Probe Definition and Characteristics*

The EndoTOFPET-US internal probe is a miniaturized PET detector head extension at the tip of an ultrasound probe meant to be placed close to the ROI. All the electronics related with the PET detector are contained in the probe flex-PCB.

Fig. 7 illustrate an overview of the main components of the probe Flex-PCB:
- 2 LYSO:Ce scintillating crystal matrices
- 2 md-SiPMs
- FPGA

In its larger version (the one represented in Fig. 7), for prostate exams, it is composed of 2 LYSO:Ce scintillating crystal matrices containing 18x9 crystals each (0.73x0.73x10

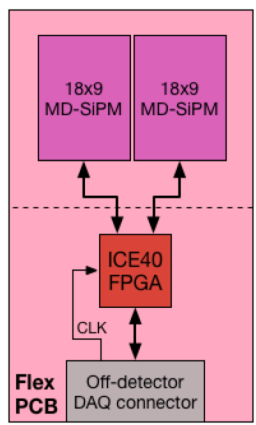

Fig. 7. Flex-PCB components. Block diagram showing the most relevant probe flex-PCB components from the DAQ point of view.

mm3) from Crystal Photonics, Inc. And in its smaller version, for pancreas exams, it comprises one of these matrices. First measurements anticipate for this scintillator matrices a LY = 12100 Ph/MeV and an energy resolution of 18%.

Electronics in the endoscopic probe requires heavy miniaturization, and thus a compact solution for the photo-detection and data processing has been implemented by the collaboration [13], which is named multi-digital Silicon Photomultiplier (md-SiPM). This device presents the same granularity as the probe crystal matrix providing a 1 to 1 crystal / photo-detector channel coupling and it offers single Single Photon Avalanche Diode (SPAD) readout, while rejecting dark counts (L0-Probe Trigger Stage), for each of its 162 channels. Within the same device, it collects and encapsulates into a digital frame all the gamma events detected during a 6.4 μs time window. For each event, it provides gamma events' time and energy information. During the characterization of the device it has been measured a single photon time resolution (SPTR) equal to 179 ps at full-width half maximum (FWHM). Further results of the characterization of the md-SiPM can be found in [14].

The digital frames produced on the md-SiPM are transmitted to a small and ultra low power ICE40 Ultra FPGA (Lattice Semiconductor, Moore Ct, Hillsboro USA). The choice of such a special FPGA comes from the fact that it has to be placed in the endoscope tip close to the md-SiPM in order to make possible a reliable communication (in terms of signal integrity) between the two devices maintaining the md-SiPM operating frequency (80 MHz).

In contrast to the FEB/D-FEB/A clock distribution, the probe flex-PCB has no jitter attenuator or low additive jitter buffers to receive the external reference clock due to the lack of space in such a small device. Nevertheless, jitter measurements have been performed on the clock signal reaching the flex-PCB, concluding that the 9.1 ps RMS jitter noise present on the system is within the md-SiPM requirements.

FPGA configuration is another interesting difference of this detector with respect to the external plate one. As it can be observed in Fig.7, the flex-PCB contains no external memory elements. In order to save the precious limited space in the probe, the FPGA model chosen has an on-chip, one-time programmable Non-volatile Configuration Memory to store configuration data. However, before the final product delivery the flex-PCB would have an external memory element connected allowing multiple reprogramming of the device.

The dashed line on Fig. 7, indicates the flexible area on the PCB that is needed for mechanical constrains on the integration with the endoscope. In the final prototype the FPGA will be mounted on a plane perpendicular to the md-SiPMs plane.

2) *Internal Probe Functionanlity*

The probe FPGA processes and filters md-SiPM data (L1-Probe Trigger Stage) on the on-detector probe electronics sending events towards the off-detector DAQ electronics with a rate up to 320 Mb/s through a LVDS pair contained in the DVI cable that connects the off-detector DAQ with the probe. This data path from probe to off-detector DAQ card is shared for the transmission of both gamma events data and system monitoring/control data, thus reducing the number of links between on-detector and off-detector DAQ segments.

The DAQ-probe communication protocol has been designed to be a scalable, lightweight and link-layer protocol whose main objectives are:
1. Move gamma events data from the probe FPGA to the DAQ across one or more serial lanes;
2. Send configuration commands from the off-detector DAQ electronics to the probe FPGA across a serial lane;

3. Monitor and control communication's parameters.

The resulting communication protocol provides a reliable communication link between probe and DAQ card thanks to data integrity features such as: error detection, DC balanced transmission, non discrete spectrum, clock recovery, data alignment, data encoding and devices synchronization. Additionally, the protocol is independent of data packet content for fast processing and flexibility.

3) *Internal Probe Validation*

Before prototyping of the final on-detector probe electronics, an intermediate setup has been prepared for the characterization of the md-SiPM itself as well as for the validation and characterization of its corresponding segment of the DAQ. This experimental test setup, shown in Fig. 8, uses the following modules:
1. The final prototype of the off-detector DAQ card;
2. A md-SiPM test board implemented for testing and characterization of the chip;
3. A ML507 Xilinx Virtex5 development kit.

Connecting boards 2 and 3 through the kit expansion headers, together they conform a complete probe equivalent. md-SiPM board has an aperture on the backside (covered by a black tissue in this picture) to allow chip characterization with laser or crystals + radioactive source. However, this particular setup has been used to perform electrical characterization and thus gamma events are triggered in the chip by an external electrical triggering pulse sent from the ML605 FEB equivalent.

In a similar way as for the case of the external plate electronics, for the internal probe electronics we have a first prototype used for chip characterization which is needed for optimum design of the final prototype electronics and which can be used at this stage for DAQ validation as well. Therefore, this setup was used to carry out experimental tests first to validate the DAQ segment tied to the internal probe and then to get link quality measures. In the latter ones we have demonstrated that, operating the link at 320 Mb/s, error free data transmission can be achieved during 24 hours continuous acquisitions. In order to perform these measurements, a specific firmware version was implemented for the off-detector DAQ card to produce md-SiPM like data towards the Xilinx Virtex5 kit that processes this data (in the same way as if it were real md-SiPM data) and send it back to the off-detector DAQ card where BER is computed.

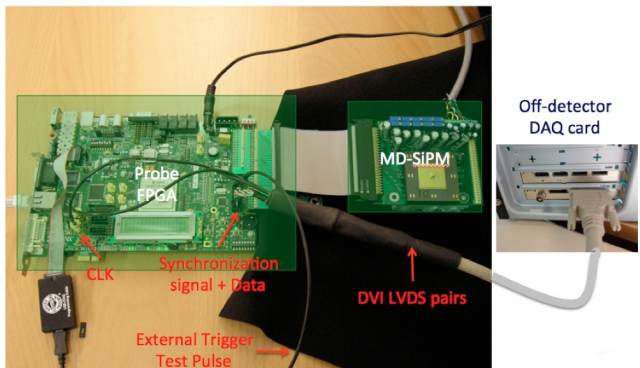

Fig. 8. Probe test setup. The right part of the picture shows the detectors IOs front panel of the Off-Detector DAQ card inside the image reconstruction workstation. It is connected through a DVI cable containing LVDS pairs to provide a CLK for the probe hardware, a synchronization signal and a data path for the readout of the md-SiPM. The picture in the left side of the figure shows the probe equivalent hardware composed of a Xilinx Virtex5 development kit (left side green box) and by a prototype of the md-SiPM (right side green box). The picture shows as well the electrical trigger link coming from the ML605 FEB equivalent kit (Fig. 6).

By means of this setup in combination with the FEB setup depicted in Fig. 6, synchronization of the two detectors with the DAQ has been tested. The procedure to test detectors synchronization by means of electrical triggering was the following:
1. Off-detector DAQ electronics sends a synchronization signal to the two detectors;
2. Each detector resets its internal counters upon reception of synchronization signal to set time 0 to the same time instant in both detectors;
3. Events are electrically triggered by means of an external test pulse, which in this setup comes from the FEB equivalent ML605 development kit;
4. Synchronization is checked by observation of a constant difference in the time stamps of the events coming from the two detectors.

C. *Off-Detector DAQ Electronics*

1) *Off-Detector DAQ Definition and Characteristics*

The off-detector DAQ electronics are responsible for the configuration, control and readout of the external plate and the internal probe. The off-detector DAQ card is implemented in a PIe enabled board integrated within the image reconstruction workstation (see Fig. 9).

Fig. 10 introduces the main components of the off-detector DAQ electronics:

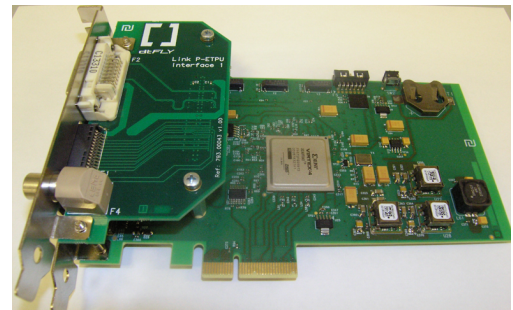

Fig. 9. Off-detector DAQ card. In the bottom of the picture the PCIe x4 connector can be easily identified. It can be seen on the left side of the picture (DAQ card front panel) how the probe IO interface is implemented as a mezzanine board allowing for the off-detector DAQ card HDMI exclusive use for future applications. The DAQ card front panel front view can be observed in Fig. 8.

- 4 HDMI connectors
- 1 DVI connector
- 1 PCIe connector
- 1 Fan-out chip
- 1 CLK buffering stage
- 1 CLK synthesizer
- 1 PROM
- 2 local oscillators
- FPGA

The board is connected to the FEB/Ds (or FEB/D chains for the daisy-chain topology) through 4 HDMI fast point-to-point links and to the endoscopic probe over a DVI connection. On each of these detectors interfaces (either for a FEB/D or for the probe) the off-detector DAQ electronics provides:
- A reference clock.
- A synchronization signal.
- A dedicated data path for the transmission of configuration commands from the off-detector DAQ to the on-detector electronics.
- A shared data path from the on-detector electronics towards the off-detector DAQ, which is shared between frequent gamma events data and spare system monitoring/control data.

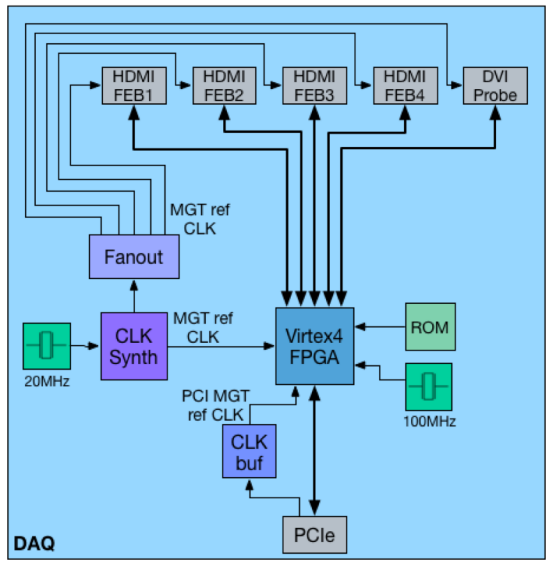

Fig. 10. Off-detector DAQ card components. Block diagram showing the most relevant off-detector DAQ card components.

The reference CLK sent towards the detectors is used by the MGTs of the detectors' FPGAs and a replica of this CLK is used in the DAQ card for its own FPGA MGTs. This reference CLK is generated by a low-jitter clock synthesizer (Integrated Circuit Systems Inc., ICS8442), it uses as source clock a 20 MHz local oscillator and it is manually configured through a switch array present on the card. These reference clocks are fanned-out towards the different system detectors through a through 1:5 low additive jitter LVDS buffers (SY89847U, Micrel, San José, USA). Thanks to this precision synthesis plus buffering, reference clock jitter is maintained low enough to even being able to avoid extra filtering upon reception on the internal probe.

The PCIe x4 connector plugs into the image reconstruction motherboard providing 4 communication lanes and a PCIe reference CLK used by the FPGA MGTs bound to the PCIe data transfer and associated logic. The reference clock is only used after being filtered through a PCIe jitter attenuator (CLK buf on Fig 10): IDT, ICS874003-02.

During power-off, the Virtex4 FPGA firmware is stored in a parallel flash memory (ROM in Fig 10): Xilinx XCF32P. On DAQ card boot-up, firmware is loaded from this memory into the FPGA.

2) *Off-Detector DAQ Functionality*

The Virtex4 FPGA is in charge of implementing the functionality required for the off-detector DAQ electronics. This FPGA is responsible for:
1. Parallel and asymmetric readout of probe and plate data;
2. Configuration of the different on-detector electronics;
3. Parallel online monitoring of the different on-detector electronics;
4. Temporal coincidence classification of the gamma events and retransmission towards the image reconstruction workstation when DAQ is set to work on coincidences operating mode;
5. Merging and temporal sorting of the gamma events coming from the different detectors and its retransmission towards the image reconstruction workstation when DAQ is set to work on singles operating mode.

For the parallel readout of the two detectors data, special considerations need to be taken into account. Not only in terms of asymmetric buffering capabilities and data rate adjustments, but also a symmetric error handling mechanism needs to be implemented in order to preserve detector uniformity as it has been explained in Section II, Level 1 triggering. In terms of asymmetric data volume flow, data generated at FEB/Ds needs special treatment for the storage of the gamma events ordered by their coarse time stamp (ordering is required for the coincidence trigger processor), otherwise the available FPGA memory resources would not be enough to handle such amount of information. Consequently, the block memory module responsible of gamma events storage from the FEB/Ds has been divided into three block memories where: the first memory stores the gamma events in the arriving order, the second memory stores the number of events occurred within the same coarse time period and the third memory stores ordered in time the addresses of the gamma events contained in the first memory. Thanks to this memory distribution approach, the required FPGA memory requirements are reduced down to a 23,68% with respect to the original requirements.

According to the philosophy of this distributed and multi-level system, the gamma events temporal coincidence classification is performed in two folds: level 2 and 3 triggering stages described in Section II. The first stage is implemented in the Virtex4 FPGA that after this filtering can procure a significantly reduced data rate towards the image reconstruction workstation.

On the other hand, the acquisition software implements the third and last triggering stage, the fine temporal coincidence triggering. The main benefit of separating coincidence filtering in two stages is that we ensure that only meaningful gamma events data will reach the reconstruction algorithms. This means that gamma events derived from Compton or optical cross-talk effects will pass the first filtering stage while totally uncorrelated gamma events will not. Afterwards, the image reconstruction software package will be able to process gamma events according to its nature while the required BW on the PCIe interface will be considerably reduced.

3) *Off-Detector DAQ Validation*

The off-detector DAQ card has been already partially validated during the on-detector validation phases explained in sections A.3 and B.3. Additionally, the PCIe interface between PC and DAQ card has been proven to provide error free data transmission at 4 Gb/s as experimental tests confirm. The procedure to acquire this quality measure was the following:
1. Off-detector DAQ card internally generates data packets with the same format as in a real detector readout scenario;
2. Generated data packets are processed and transferred to the PC via the PCIe interface;
3. PC DAQ readout software reads the generated data packets while checking data integrity and computing readout rate.

## IV. OPERATING MODES

Once every component of the system has been validated, during the operation of the scanner, the procedure depicted in Fig. 11 is conducted:

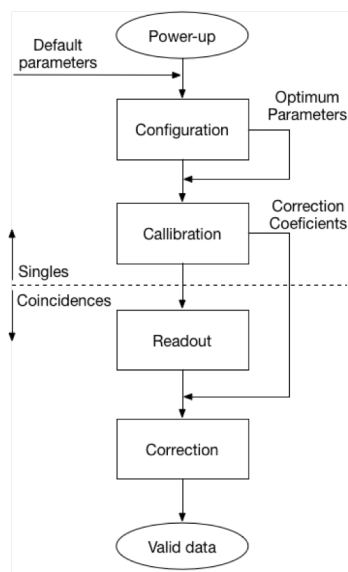

Fig. 11. Scanner operating modes. Block diagram illustrating the sequential actions performed during the scanner operation.

There are two set of parameters needed to optimize the system performance: the configuration parameters and the correction coefficients. The system characteristics vary from pixel to pixel; therefore, these sets of parameters are channel-specific.

The configuration parameters are internal parameters of the ASICs and md-SiPM. For the external plate case, they specify the amplification gain of the signals produced by the photo-detectors as well as the thresholds used in time and energy measurements [8-9]. For the internal probe, this set of parameters specifies key characteristics such as the channel masking profile (DCR reduction), smart reset frequency (dead time reduction), or time and energy thresholds [13].

On system start-up the default parameters, are loaded during the configuration phase. Then, a scan around the default parameters is performed to obtain the optimum configuration parameters.

Once the optimum configuration parameters are loaded on the ASICs and md-SiPM, the calibration stage is run in order to extract the correction coefficients that need to be applied to correct the Time to Digital Converters (TDC) non-linearities and the energy measure.

There are two independent procedures for time and energy calibration. Time can be calibrated via external electrical triggering of the device, whereas energy requires the use of different radioactive sources to extract its calibration curve.

It should be noticed, that during the initial configuration and calibration phases the system would work in singles mode, i.e. every single gamma event triggering the system would be read and transferred to the acquisition software. On the other hand, once the system is optimized and correction factors are extracted, during data tacking the system would work on coincidences mode, i.e. only the time coincident events (based on the coarse L2 triggering policy: 12.5 ns coincidence time window) would be transferred and corrected.

## V. THROUGHPUT MEASUREMENTS

In order to analyze the DAQ performance in terms of data throughput the following setup has been used: two FEB/A detector modules mounted on one FEB/D super-module, which is read by the off-detector DAQ card. The data has been triggered on the FEB/A ASICs by means of the electrical trigger test-pulse, which is launched by the FEB/D FPGA. The number of active channels in the ASICs has been progressively activated and the test-pulse frequency has been adjusted to produce a data rate sweep over the complete ASICs available BW. The DAQ input data rate has been computed as the amount of events per time electrically triggered on the ASICs and the output rate as the amount of events written on disk by the DAQ card. Fig. 12 shows the results of the throughput analysis for one and two FEB/As.

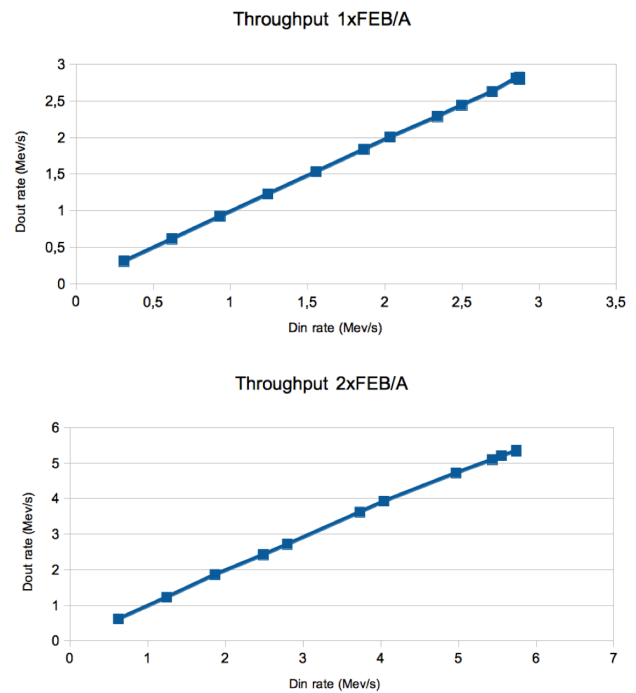

Fig. 12. DAQ throughput measurements for one and two FEB/As connected on a FEB/D and read by the DAQ card.

The current FEB/A configuration is capable to produce a maximum data rate of 2.87 Mevents/s and, as it can be noticed in Fig. 12, the DAQ system is capable of handling this data rate. The FEB/D to DAQ link has been tested successfully up to 5.74 M ev/s. Small differences in the input and output rates can be appreciated and it can be seen how the difference grows with the number of FEB/As. These disparities are due to the increment of the packet overhead as the packet goes through the system.

Taking into account the available BW in all the system interfaces and current results for one and two FEB/As, it is a reasonable assumption that these results would scale linearly when more FEB/As are read by the DAQ up to a maximum of almost 23 Mevents/s in the scenario of a FEB/D fully populated with 8 FEB/As. According to the expected system gamma event rates anticipated in Section II, where the external plate is assumed to produce a total of 40 Mevents/s (i.e. 10 Mevents/s per FEB/D), these tests already foresee the compliance of the DAQ system with the rates expected in singles mode.

## VI. CONCLUSIONS

An Asymmetric Data Acquisition system specific for an endoscopic PET-US scanner has been designed, implemented and tested. The system is capable to configure and readout two different detectors, which lead to an asymmetric readout scenario.

The presented DAQ is distributed and balanced across the different system components, thus providing the opportunity to implement a multi-level triggering scheme that allows fine central adjustment and progressively reduces data rate. This processing approach prevents the system from filtering out meaningful data which is highly precious in this scanner scenario where FoV is limited and sensitivity is low.

Due to the design intrinsic system flexibility/expandability, this DAQ could be easily scalable and its compatibility on future applications (such as small animals scanners or particle therapy online radiation monitoring) is guaranteed.

As a proof-of-concept, the EndoTOFPET-US DAQ system has been successfully validated by means of detectors and DAQ card prototypes leading us to the conclusion that a distributed and multi-level triggering DAQ system is suitable for endoscopic PET detectors. For the next phase of this project, a complete prototype with all the final electronics for the two on-detector sections will be used for the system commissioning. The initial phase of the project required independent work of the different partners having major interaction within work packages but little communication between them. The final phase, require major interaction between all groups in order to put together the complete scanner.

Once the complete scanner is available, in the context of DAQ system characterization, the coherence between current results and the exposed expectations on scalability will be checked and a study on the benefits of the multi-level triggering scheme will be analyzed.